# Lithographically fabricated magnifying Maxwell fisheye lenses

Vera Smolyaninova,[1] William Zimmerman,[1] Christopher Jensen,[1] Anthony Johnson,[1] David Schaefer,[1] Igor Smolyaninov[2,*]

[1]Department of Physics, Astronomy and Geosciences, Towson University, 8000 York Rd., Towson, MD 21252, USA
[2]Department of Electrical and Computer Engineering, University of Maryland, College Park, MD 20742, USA
*Corresponding author: smoly@umd.edu

**Recently suggested transformation optics-based magnifying Maxwell fisheye lenses, which are made of two half-lenses of different radii, has been fabricated and characterized. The lens action is based on control of polarization-dependent effective refractive index in a lithographically formed tapered waveguide. We have studied wavelength and polarization dependent performance of the lenses.**

Transformation optics (TO) has recently become a useful methodology for the design of unusual optical devices, such as novel metamaterial lenses and invisibility cloaks. Unfortunately, typical TO designs require metamaterials with low-loss, broadband performance, which appear difficult to develop. These difficulties are especially severe in the visible frequency range where good magnetic performance is limited. On the other hand, very recently we have demonstrated that many transformation optics and metamaterial-based devices requiring anisotropic dielectric permittivity and magnetic permeability may be emulated by specially designed tapered waveguides [1]. This approach leads to low-loss, broadband performance in the visible frequency range, which is difficult to achieve by other means. We have applied this technique to broadband electromagnetic cloaking in the visible range [1] and successfully extended it to birefrigent TO devices, which perform useful and different functions for mutually orthogonal polarization states of light [2]. In this Letter we have applied this approach to lithographically fabricated magnifying Maxwell fisheye lenses, which were originally introduced in a microdroplet form [3].

Unlike the earlier microdroplet design, which is difficult to fabricate and control, our current design is based on lithographically defined metal/dielectric waveguides. Adiabatic variations of the waveguide shape enable control of the effective refractive indices experienced by the TE and TM modes propagating inside the waveguides, as shown in Fig. 1(a). They are defined as $n_{eff} = k\omega/c$ for each polarization, where the $k$ vector at a given frequency $\omega$ is calculated via the boundary conditions at two interfaces as follows:

$$\left(\frac{k_1}{\varepsilon_m} - \frac{ik_2}{\varepsilon}\right)\left(k_3 - \frac{ik_2}{\varepsilon}\right)e^{-ik_2 d} = \left(\frac{k_1}{\varepsilon_m} + \frac{ik_2}{\varepsilon}\right)\left(k_3 + \frac{ik_2}{\varepsilon}\right)e^{ik_2 d} \quad (1)$$

for the TM, and

$$(k_1 - ik_2)(k_3 - ik_2)e^{-ik_2 d} = (k_1 + ik_2)(k_3 + ik_2)e^{ik_2 d} \quad (2)$$

for the TE polarized guided modes, where the vertical components of the wavevector $k_i$ are defined as:

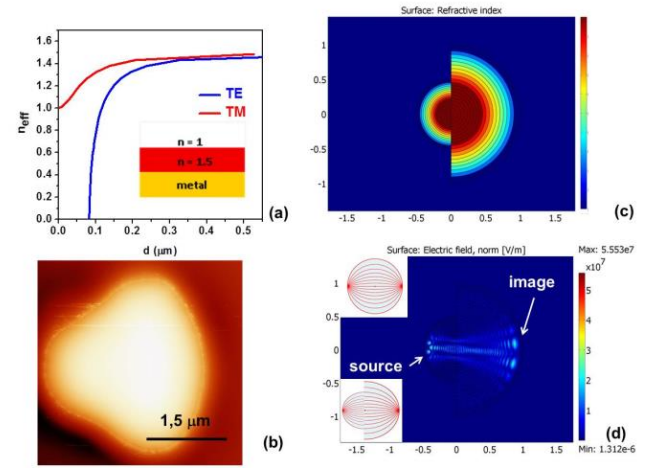

Fig. 1. (a) Effective refractive index plotted as a function of thickness of a tapered waveguide. The waveguide geometry is shown in the inset. (b) AFM image of a lithographically defined individual magnifying fisheye lens made of two half-lenses of different radii. (c) Corresponding spatial distribution of the effective refractive index. (d) COMSOL Multiphysics simulation of the fisheye lens image magnification. The insets illustrate ray propagation in the original and the magnifying fisheye lenses.

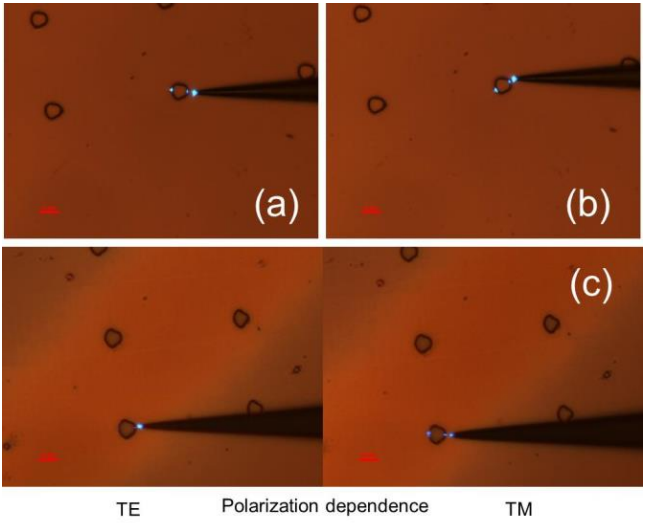

Fig. 2. Experimental testing of angular (a,b) and polarization (c) performance of the magnifying fisheye lenses at λ=488 nm. The scale bar length is 5 μm in all images.

$$k_1 = \left(k^2 - \varepsilon_m \frac{\omega^2}{c^2}\right)^{1/2} \quad (3)$$

$$k_2 = \left(\frac{\omega^2}{c^2}\varepsilon - k^2\right)^{1/2} \quad (4)$$

$$k_3 = \left(k^2 - \frac{\omega^2}{c^2}\right)^{1/2} \quad (5)$$

in metal, dielectric, and air, respectively. As illustrated in Fig. 1(a), effective birefringence for the lowest guided TM and TE modes appears to be very strong at waveguide thickness $d<0.4$ μm, and both polarizations demonstrate strong index dependence on the waveguide thickness. This behavior may be used in building non-trivial birefringent TO devices if a waveguide thickness as a function of spatial coordinates $d(r)$ may be controlled lithographically with enough precision.

We have developed a lithography technique which enables such $d(r)$ shape control of the dielectric photoresist on gold film substrate, as illustrated in Fig. 1(b). Shieply S1811 photoresist having refractive index $n \sim 1.5$ was used for device fabrication. In traditional lithographic applications for the best results of pattern transfer the profile of resist should be rectangular or even with overhang. Our purpose is different. We want to create a more gradual edge profile. This can be done by disregarding typical precautions employed to make the edges sharp. Instead of contact printing (when mask is touching the substrate), we used soft contact mode (with the gap between the mask and the substrate). This allows for the gradient of exposure due to the diffraction at the edges, which leads to a gradual change of thickness of the developed photoresist. Underexposure and underdevelopment were also used to produce softer edges. This technique has been used previously to fabricate such TO-based devices as a modified Luneburg lens [2]. However, no image magnification has been demonstrated in these experiments. On the other hand, as was noted in [3], it is relatively straightforward to incorporate image magnification into such TO lens designs as Eaton and Maxwell fisheye lenses.

The refractive index distribution in a Maxwell fisheye lens is defined as

$$n = 2n_1\left(1 + \frac{r^2}{R^2}\right)^{-1} \quad (6)$$

at $r < R$, where $2n_1$ is the refractive index at the center of the lens, and R is the scale factor. A reflective surface is assumed to be placed at $r = R$, so that $n < 1$ values of the refractive index do not need to be used. In our experiments the role of such reflective surface is played by the lens edge. On the other hand, an inverted Eaton lens [5] is defined as $n = n_1$ for $r < R$, and

$$n = n_1\left(\frac{2R}{r} - 1\right)^{1/2} \quad (7)$$

for $r>R$. Since the refractive index distribution in the fisheye lens is obtained via the stereographic projection of a sphere onto a plane [6], points near the lens edge correspond to points located near the equator of the sphere. Therefore, these points are imaged into points located near the opposite lens edge, as shown in the inset in Fig. 1(d). The inverted Eaton lens has similar imaging properties.

As demonstrated in [3], both refractive index distributions may be emulated with a suitable $d(r)$ profile of a thin dielectric placed on top of a metal film, which is also evident from Fig. 1(a). As shown in Fig. 1(c,d), two halves of either Maxwell fisheye or inverted Eaton lens having different values of parameter R may be brought together to achieve image magnification. The image magnification in this case is $M = R_1/R_2$. Our numerical simulations in the case of $M = 2$ are presented. Since the sides of the lens play no role in imaging, the overall shape of the imaging device can be altered to smooth the sharp corners, resulting in the magnifying fisheye lens shape shown in Fig. 1(b), which was fabricated using the lithographic technique described above.

Experimental images in Fig. 2 demonstrate measured performance of the designed magnifying fisheye lenses. In these experiments a near-field scanning optical microscope (NSOM) fiber tip was brought in close proximity to the arrays of lithographically formed TO devices and used as an illumination source. As expected from the numerical simulations, an image of the NSOM tip was easy to observe at the opposite edge of the lens. Angular and polarization testing of individual lenses in the array agrees well with theoretical modelling presented in Fig. 1(c,d). As illustrated by Fig. 1(a), the same $d(r)$ profile produces a different refractive index distribution for TM and TE polarized light.

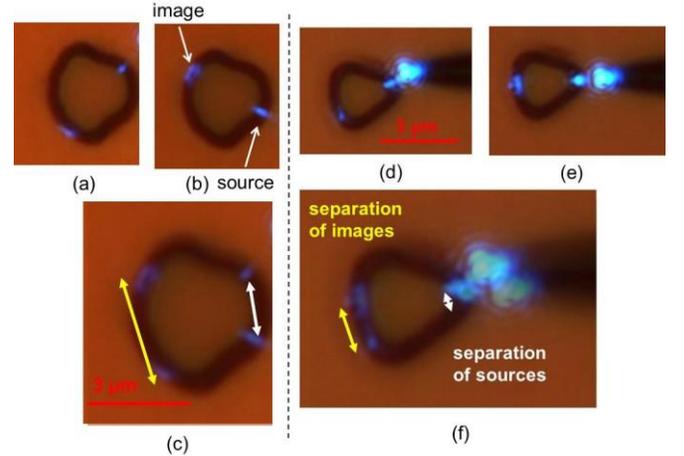

Fig. 3. Experimental testing of image magnification at $\lambda = 488$ nm of two fisheye lenses with different $M = R_1/R_2$ ratio: (a,b) Original magnified images obtained at different source positions. The location of image and source are indicated by the arrows. (c) Digital overlap of the images in (a) and (b) indicates that image magnification is close to the design value $M = 2$. (d,e) Similar original images and (f) the digital overlap image obtained with a different magnifying lens designed for $M = 3$.

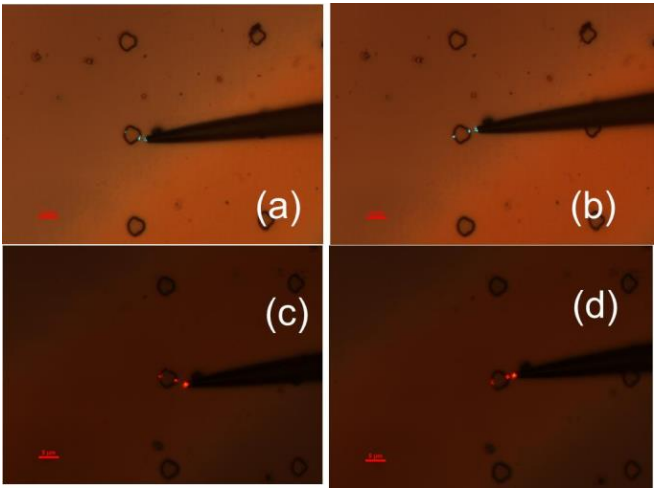

Fig. 4. Experimental verification of broadband performance of the magnifying Maxwell fisheye lens. (a,b) Images taken at $\lambda$ = 515 nm. (c,d) Images taken at $\lambda$ = 633 nm. The scale bar length is 5 μm in all images.

Due to near zero effective refractive index near the device edge, a fisheye lens for TM light will operate as a spatial (directional) filter for TE light [2]. We should note that while the NSOM fiber tip emits unpolarized light, polarization response of the image produced by a TO device can be clearly separated into the TM and TE contributions with respect to the plane of incidence of the source light. Polarization behavior of lenses in Fig. 2 demonstrates excellent agreement with theory.

Image resolution and magnification of the fabricated magnifying Maxwell fisheye lenses may be evaluated based on images of lens testing presented in Fig. 3. Experimental testing of two fisheye lenses having different $M = R_1/R_2$ ratio is shown in these images. The image resolution appears to be close to diffraction-limited (~$0.6\lambda$ at 488 nm), while image magnification is close to the design values $M$ = 2 and $M$ = 3, respectively. We should also note a very broad (almost 180°) angular range of the magnifying fisheye lens operation.

The projected broadband performance of the magnifying Maxwell fisheye lenses has been verified in the $\lambda$ = 488 – 633 nm range. Examples of such testing at 515 nm and 633 nm are presented in Fig. 4 (compare these images with Fig. 2(a,b) obtained at 488 nm). We have also verified (see Fig. 5) that the same lens used in reverse direction may be utilized to achieve image reduction. This fact is not trivial since the lens geometry is obtained by "gluing together" two halves of the Maxwell fisheye lenses having considerably different radii, which from the ray optics point of view may lead to ray scattering by the lens edges. Potentially, such an arrangement of the magnifying fisheye lens may find lithographic applications.

In conclusion, we have reported the first experimental realization of TO-based birefringent lithographically fabricated magnifying Maxwell fisheye lenses, which operate over a very broad angular range. The lens action is based on control of polarization-dependent effective refractive index in a tapered waveguide. We have studied wavelength and polarization dependent performance of the lenses. The fabricated

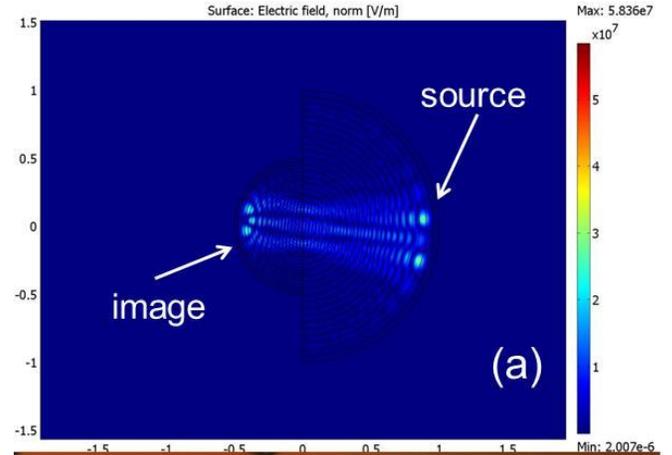
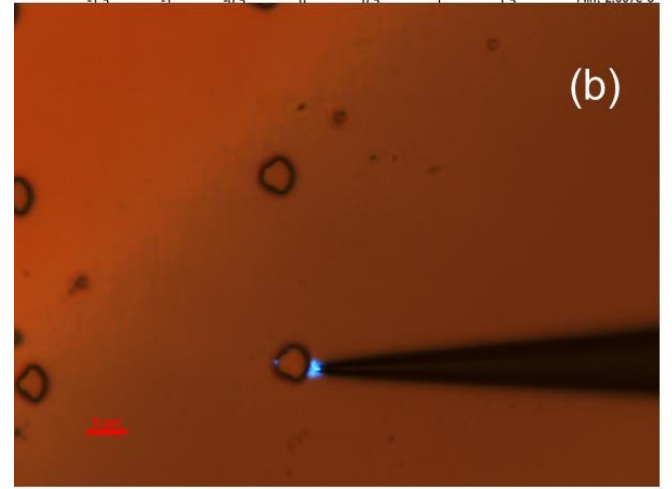
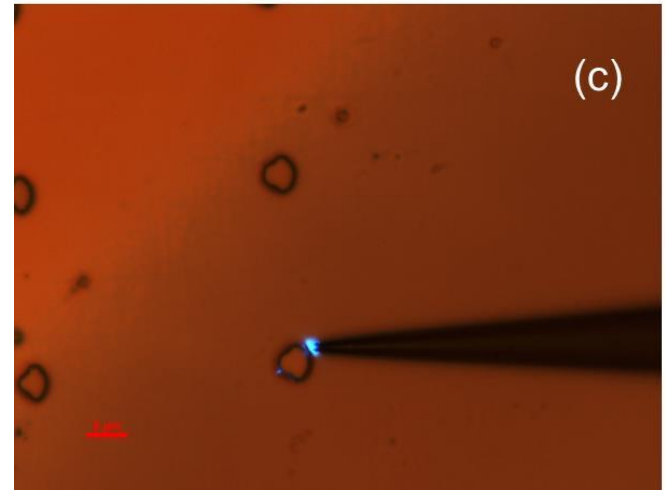

Fig. 5. Operation of the magnifying Maxwell fisheye lens in reverse direction, which leads to image reduction. (a) COMSOL Multiphysics simulation of the fisheye lens image reduction using refractive index distribution corresponding to the experimental variation of the waveguide thickness. (b,c) Experimental testing of angular performance of the fisheye lens used in reverse direction. $\lambda$ = 488 nm. Image reduction factor $M$ = 1/2 is observed. The scale bar length is 5 μm in all images.

TO designs are broadband, which was verified in the 488-633 nm wavelength range. Our technique opens up an additional degree of freedom in optical design and considerably improves our ability to manipulate light on submicrometer scale.

**Funding.** This research was supported by the National Science Foundation (NSF) grant DMR-1104676.

**Acknowledgment**. We are grateful to Jeff Klupt for experimental help.